\documentclass[12pt]{article}
\usepackage{graphicx}
\usepackage{url}
\usepackage{amsmath,amssymb,amsfonts,geometry}
\usepackage{amsthm}
\usepackage{float,color}
\usepackage{graphics}
\title{Adiabatic quantum computing solution of the knapsack problem} %with integer weights and profits 
\author{Mark W. Coffey\\
Department of Physics\\
Colorado School of Mines\\
Golden, CO  80401\\
(Received $\mbox{~~~~~~~~~~~~~~~~~~~~~~~~~~~~~~~2017}$)}
\date{January 17, 2017}
\newcommand{\braket}[2]{\langle #1|#2\rangle}
\begin{document}
\maketitle
%\vspace{.25cm}
\baselineskip=25 pt
\begin{abstract}

We illustrate the adiabatic quantum computing solution of the knapsack problem with both integer profits and
weights.  For problems with $n$ objects (or items) and integer capacity $c$, we give specific examples using both an Ising class problem Hamiltonian requiring $n+c$ qubits and a much more efficient one using $n+[\log_2 c]+1$ qubits.  The discussion includes a brief mention of classical algorithms for knapsack,
applications of this commonly occurring problem, and the relevance of further studies both theoretically
and numerically of the behavior of the energy gap. % mention the bit on qntm search in here too?
Included too is a demonstration and commentary on a version of quantum search using a certain Ising model.
Furthermore, an Appendix presents analytic results concerning the boundary for the easy-versus-hard
problem-instance phase transition for the special case subset sum problem.

\end{abstract}
 
%\vspace{.20cm}
\medskip
\baselineskip=15pt
\centerline{\bf Key words and phrases}
\medskip 

\noindent
adiabatic quantum computing, knapsack with integer weights, spectral gap, binary variable, Ising
Hamiltonian, subset sum, phase transition, quantum search                   % quantum search?

%\vspace{.25cm}
\vfill
\centerline{\bf 2010 PACS codes}
03.67Ac, 03.67.Lx % for qntm algos, qntm comp. implementations, respectively.

%\vspace{.25cm}
%\vfill
%\centerline{\bf 2010 AMS codes} --TBS or remove
%33C05, 33C20, 05A10, 81P45

\baselineskip=25pt
\pagebreak
\medskip
\centerline{\bf Introduction}
\medskip

Adiabatic quantum computing (AQC) is an approach (polynomially) equivalent to the circuit model of quantum computing \cite{doritetal}.  It has the attraction of being suitable for problems of the combinatorial
optimization sort including partitioning, covering, traversing trees and graphs, and logical satisfiability.  In fact, early important papers in this field were concerned with the latter topic \cite{farhi1,farhi2}.
% cite too an early E. Farhi paper with tree traversal?
This paper concentrates on the AQC solution of the knapsack apportioning problem, but we also remark
on such solution of other problems.  % including the 'simple' search problem in the Conclusion section of
% A. Lucas' paper       
% there is a Nov. 2016 review of aqc by Albash and Lidar on the arXiv, but it does not seem to cite
% A. Lucas's paper--email them about it? and possibly my ms?

Within the AQC method, the ground state of a problem Hamiltonian encodes the solution of interest.
The hardware solution arises from the slow-enough evolution of the ground state of an initial
simpler Hamiltonian to that of the problem Hamiltonian.  Then the problem solution may be read out.
Without going in to further detail, the required evolution time is dictated by the inverse square of
the spectral gap, the minimum energy difference between the ground and first excited states. % ref(s) here?
Traditionally the adiabatic theorem has been based upon time-dependent perturbation theory, and of course
more recently there is a variety of more specific and rigorous results.
Experimentally, so far AQC has been demonstrated in NMR (nuclear magnetic resonance) and Josephson-junction-based systems.
% for NMR impl. see Mitra et al. (2005) and there is likely more recent exptl work. 

% can compare the intro of the qntm annealing for planning article of Rieffel et al.:
Let $T$ be a sufficiently large evolution time, as predicated by the gap and the magnitude of the matrix elements $\braket{n}{{{\partial H} \over {\partial t}}|k}$, and let the normalized time be $s=t/T$.  Then the total Hamiltonian
takes the form $H(s)=a(s)H_0+b(s)H_p$, where the functions $a$ and $b$ are monotonically 
decreasing and increasing, respectively, and $a(0)=1$, $a(1)=0$, and $b(0)=0$ and $b(1)=1$.  A key 
requirement is that the initial $H_0$ and problem $H_p$ Hamiltonians do not commute, lest the gap may vanish.
There is an infinite number of choices of the functions $a$ and $b$, and, furthermore, a full
Hamiltonian such as $H(s)=a(s)H_0+c(s)H_a+b(s)H_p$ with $c(0)=c(1)=0$ with additional term $H_a$ is
certainly possible.  For the purpose of concreteness in implementation we will later restrict to $H(s)=(1-s)H_0+sH_p$. %for the remainder of this article.  
However, it may be noted that using such a linear interpolation does not provide a computational
advantage for quantum search, and that a rescaling is required in order to obtain the optimality of
Grover's algorithm \cite{roland}.

A convenient, restricted, and certainly not universal class of Hamiltonians is the Ising model
$$H=-\sum_i h_i\sigma_z^i-\sum_{i<j}J_{ij}\sigma_z^i\sigma_z^j, \eqno(1) $$
where $h_i$ correspond to magnetic field strengths and $J_{ij}$ to spin-spin couplings.  In particular,
there are no transverse field contributions to this class of problem Hamiltonians, and it is 2-local:
there are no interactions of $3$ or more spins.  The model (1) conveniently extends semiclassical
magnetic models with spins $s_i=\pm 1$ to the quantum domain with operators $\sigma_z^i=$diag$(1,-1)$ 
acting on the $i$th qubit.

Again there is much choice in the initial Hamiltonian $H_0$.  One such that is very convenient,
but not necessarily providing the best behavior of the spectral gap, is $H_0=-h_0\sum_i \sigma_x^i$,
% get into D Wave in the Intro?
where $\sigma_x^i$ is the NOT gate on the $i$th qubit.  Among many others, $H_0=-h_0\sum_{i<j} \sigma_x^i
\sigma_x^j$ is an alternative. % also possible. % or write:  Among many other operators ...
% cite the McGeogh etc. monograph in the Intro?

In the following sections, we set up and illustrate the knapsack problem, commenting on classical
algorithms, then describe high-level implementations for Ising models for an AQC solution.  
We also discuss the AQC solution of some other, more restricted problems. %i.e., subset sum, the search
% problem at the end of the Lucas paper. % the word 'restricted' is used pretty often in this Intro.

Concerning NP-difficult combinatorial optimization problems, these require an exponentially large amount
of at least one resource for solution in the worst case, and the AQC methodology will not always provide an
advantage \cite{vandam}. % cite an early Farhi paper here too? %cite Ben Reichardt paper here too?
With $N$ a measure of problem size, the computational cost may then vary as O$[\exp(\alpha N^\beta)]$
for very large $N$ and positive $\alpha$ and $\beta$.
When the AQC approach is more effective than classical algorithms, we expect as a result the
exponent $\alpha N^\beta$ to typically be reduced to $\alpha N^\beta/2$.  While a further reduction in computational
cost would be appealing, this is still very significant for practical-size problems.
% as for Lucas Ref. [31] for Ramsey numbers?

\medskip
\centerline{\bf The Knapsack problem}
\medskip

% see my .ppt charts and Kellerer et al. book--

We will be concerned with knapsack instances with integer weights $w_j$ and profits $p_j$.
The input for the knapsack problem consists of these numbers together with a capacity $c$, also
taken to be an integer, and $n$, the number of items.  Formulated as a binary programming problem,
knapsack then is comprised as follows.
$$(KP) ~\mbox{maximize} ~~\sum_{j=1}^n p_j x_j$$
$$ ~~~~~~~~~~~~~~~\mbox{subject to} ~~\sum_{j=1}^n w_j x_j \leq c,$$
with $x_j\in\{0,1\}$ for $1 \leq j \leq n$.
The knapsack problem and its various extensions have numerous applications in packing and stock cutting
problems, financial decision making, asset-backed securitization, and combinatorial auctions \cite{kellerer}.
In the latter area, a bundle of goods is sold, not just a single item.  Moreover, solutions of knapsack
may serve to find a solution of a more complicated problem, which could include scheduling.

The special case of $p_j=w_j$ for $j=1,\ldots,n$ is referred to as the subset sum problem.
It is still NP difficult.  Although many algorithms for knapsack are suitable for it, it may also
be attacked by more specialized means.  % what would be some detail for this last statement?
% attacked->tackled or some other word?

Some remarks on the input for the knapsack problem are in order, beyond that we take $n\geq 2$.
For each weight we require $w_j \leq c$, and for their sum $\sum_{j=1}^n w_j > c$.  If the latter
condition did not hold, we would simply take $x_j=1$ for $j=1,\ldots,n$.  In addition, without loss
of generality we may assume $p_j>0$ and $w_j>0$ for $j=1,\ldots,n$.  If otherwise a value $p_k$ or
$w_k$ is negative, the problem instance may be manipulated in order to have positive weights and profits.

% probably should move this paragraph to after the problem statement/formulation:
For the knapsack problem, as with other NP difficult problems, we expect that there is one or
more phase transitions in problem difficulty.  % maybe comment later on this as regards the subset sum problem
We may expect that in some sense average problem instances are easy and only require polynomial
amounts of computational resources,  but that certain subsets, as with weights and/or costs with a 
large least common multiple require exponential amounts of resource. % [or is it the GCD?]
For the easier number partitioning problem, the question of a phase transition has been fairly well
studied and characterized \cite{gent,gent2,mertens,sasamoto,sasamoto2}.  If for the knapsack problem the weights are
drawn from $\{1,2,\ldots,L_w\}$ and the profits from $\{1,2,\ldots,L_p\}$, then two of the parameters 
describing the phase transition(s) may be $\kappa_w=\log_2 L_w/n$ and $\kappa_p=\log_2 L_p/n$, yet the
capacity must also be brought in.  Section 4.3 of the review \cite{smith} may be consulted for a 
discussion of instance difficulty for knapsack.  In the Appendix we include analytic results which
complement the analysis of hard versus easy cases of a subset sum problem.  

% do searches on both:  knapsack and number theory
%                       phase transition (and/or stat. mech.) and knapsack
% have a mention too of the various financial, etc. applications/variations of knapsack--done above.
As an illustration of knapsack, and which will also serve as a test case in the next section, we
consider an instance with $n=7$ and $c=9$:
\begin{center}
\begin{tabular}{|c|ccccccc|} \hline
$j$   & 1 & 2 & 3 & 4 & 5 & 6 & 7 \\ \hline
$p_j$ & 6 & 5 & 8 & 9 & 6 & 7 & 3 \\ 
$w_j$ & 2 & 3 & 6 & 7 & 5 & 8 & 4 \\ \hline
\end{tabular}
\end{center}
The items have been listed in decreasing order of efficiency ratio $r_j=p_j/w_j$.  Accordingly,
greedy algorithms will return a solution of either items 1 and 2 with profit 11 or items 1, 2, and 7
with profit 14.  However, the optimal solution with profit value 15 comes from items 1 and 4.

In addition to greedy algorithms and branch-and-bound, a classical algorithm that may be used to
solve knapsack is dynamic programming (DP) by weights or by profits.  When using DP by weights, a
series of problems for capacity values $d=0,\ldots,c$ computes profit values $z_j(d)$.  Then a
certain recursion is applied for $j=1$ up to $j=n$, yielding the optimum solution value as $z_n(c)$.
I.e., the all-capacities knapsack problem is solved in this procedure.  With returning the optimal
profit value $z^*$ but not explicitly the optimal solution set of items, the running time is $O(nc)$.
% see esp. pp. 20-23 of Kellerer et al.

\medskip
\centerline{\bf AQC solution via Ising models}
\medskip

We present two versions of a problem Hamiltonian for the knapsack problem as formulated in \cite{lucas}.
The second form is implicit in \cite{lucas} so that we provide more explanation for it.  We have
implemented both of these Hamiltonians in Mathematica$^\copyright$ together with initial Hamiltonians
in order to produce a simulation of the spectral gap.

Firstly let $x_i$ be binary variables for $1\leq i\leq n$ and $y_k$ be binary variables for $1\leq k\leq c$.
A problem Hamiltonian may be written as $H=H_A+H_B$, where
$$H_A=A\left(1-\sum_{j=1}^c y_j\right)^2+A\left(\sum_{j=1}^c jy_j-\sum_{i=1}^n w_i x_i\right)^2, \eqno(2a)$$
and
$$H_B=-B\sum_{i=1}^n p_ix_i. \eqno(2b)$$
The term $H_B$ serves to maximize the profit and the term $H_A$ ensures that the total weight constraint
is satisfied.  In this set up, only a single $y_j$ variable will be nonzero.  The condition on $A$ and $B$ is
$0<B \mbox{max}(p_j)<A$.  The number of qubits required is $n+c$, the sum of the number of items and the
capacity.  In the quantum version the binary variables $x_i$ are represented by the operators
$(I+\sigma_z^i)/2=$diag$(1,0)$.

We now make it explicit how the number of needed qubits may be substantially reduced to $n+[\log_2 c]+1$.
Let the integer $M$ be determined from $M=[\log_2 c]$, thus ensuring that $2^M\leq c<2^{M+1}$.
The reduced problem Hamiltonian then consists still of $H_B$, with now
$$H_A=A\left(\sum_{j=0}^{M-1} 2^jy_j+(c+1-2^M)y_M-\sum_{i=1}^n w_i x_i\right)^2. \eqno(3)$$
There are now $M+1$ new binary variables $y_0,y_1,\ldots,y_M$, and generally several of them will be
nonzero together.  This occurs due to nonuniqueness to represent a total weight.

As an example suppose that the capacity $c=10$.  We may quickly verify that all total weights up to $c$
may be represented with $4$ binary variables.  Here $M=3$ and the combination
$y_0+2y_1+4y_2+3y_3$ may represent all values $1,2,\ldots,10$.  For instance, total weight 6
results for either $y_1=1=y_2$ and $y_0=0=y_3$ or $y_0=1=y_1=y_3$ and $y_2=0$.  Hence degenerate ground
state solutions are now possible.

% next describe AQC simulation for the above test case (if not also for another) and refer to first figure.
We have principally used $H_0=-h_0\sum_i \sigma_x^i$ as the initial Hamiltonian in the simulations,
where in the first version $i$ runs from $1$ to $n+c$, and in the second from $1$ to $n+M+1$.
Since the problem Hamiltonians are diagonal by definition, implementations with programming languages which
are efficient with lists can benefit from this aspect.  Comparing the Ising form (1) with (2) and (3) we
see that the profits $p_j$ and weights $w_j$ enter the magnetic field parameters $h_i$ and that the
product of weights $w_iw_j$ contributes to the qubit-qubit couplings $J_{ij}$.

Figures 1a and 1b show respectively the ground state energy, first excited state energy, and their
difference, and the latter difference separately for the knapsack instance with $n=5$ and $c=7$:
\begin{center}
\begin{tabular}{|c|ccccc|} \hline
$j$   & 1 & 2 & 3 & 4 & 5 \\ \hline
$p_j$ & 8 & 3 & 5 & 6 & 9 \\ 
$w_j$ & 1 & 2 & 1 & 3 & 2 \\ \hline
\end{tabular}
\end{center}
The optimal solution consists of items 1, 3, 4, and 5, with profit 28 and total weight 7 matching the
capacity.  The adiabatic solution is based upon using problem Hamiltonian (2) so that 12 qubits are needed.
A fairly common feature with this approach is a small gap at smaller values of $s$, followd by a very
nearly linear behavior.

\begin{figure}[h]
\begin{center}
\includegraphics[height=2.5in,width=4.0in,angle=0]{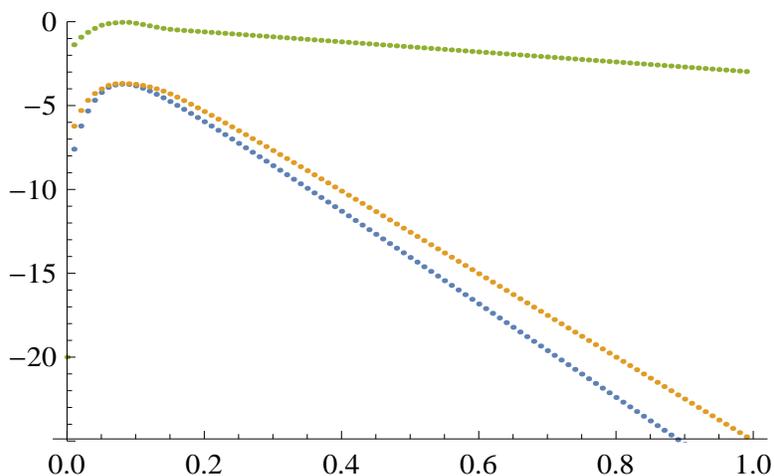} %vs .eps
\caption{Fig. 1a. Ground and first excited state energies, and their difference, vs $s=t/T$.}
\end{center}
\end{figure}

\begin{figure}[h]
\begin{center}
\includegraphics[height=2.5in,width=4.0in,angle=0]{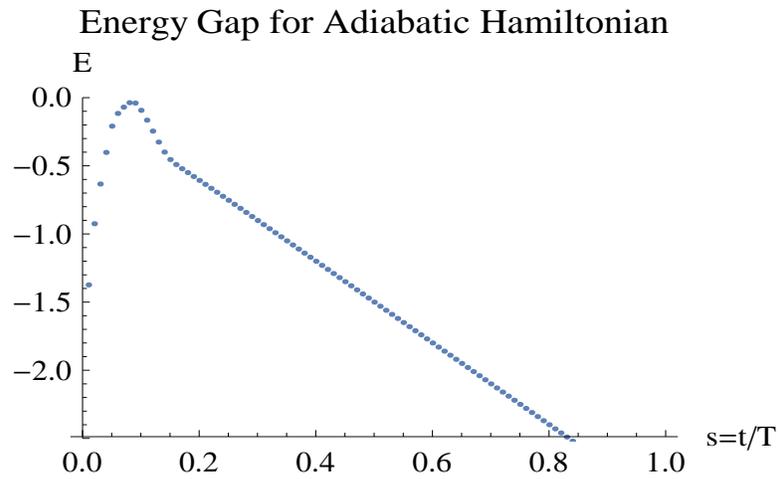} %vs .eps
\caption{Fig. 1b. Energy difference of the first and ground states energies vs $s=t/T$.}
\end{center}
\end{figure}

An example for the gap of the test problem of the previous section is given in Figs.\ 2a and 2b.
Here the problem Hamiltonian is based upon (2b) and (3).  
% to describe feature(s) of it, ....................
% may end up stating that there is typically a small gap for smaller s, but it is sometimes too almost
% near s = 1/2.
Again the energy difference between the ground and first excited states is smallest for the earlier
times.  This seems to be a fairly generic feature, but some problem instances have the smallest such
difference at significantly larger values of $s$.

\begin{figure}[h]
\begin{center}
\includegraphics[height=2.75in,width=4.25in,angle=0]{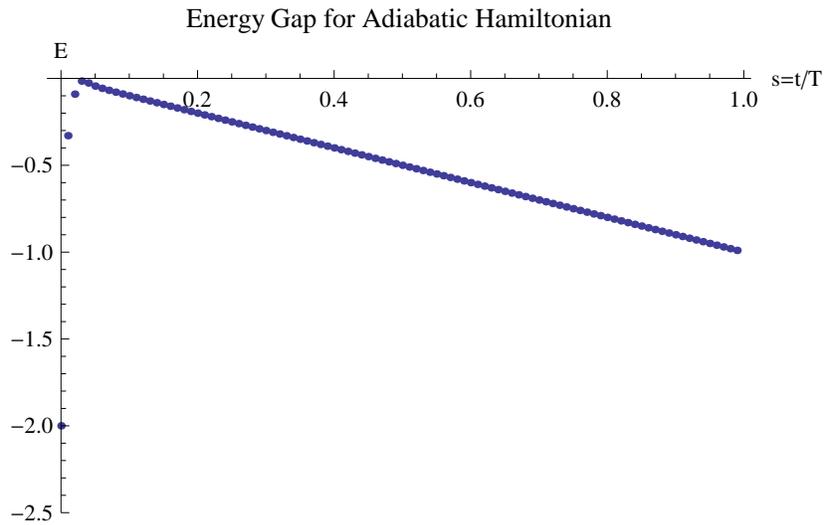} %vs .eps
\caption{Fig. 2a. Energy difference of the first and ground states energies vs $s=t/T$.}
\end{center}
\end{figure}

\begin{figure}[h]
\begin{center}
\includegraphics[height=2.75in,width=4.25in,angle=0]{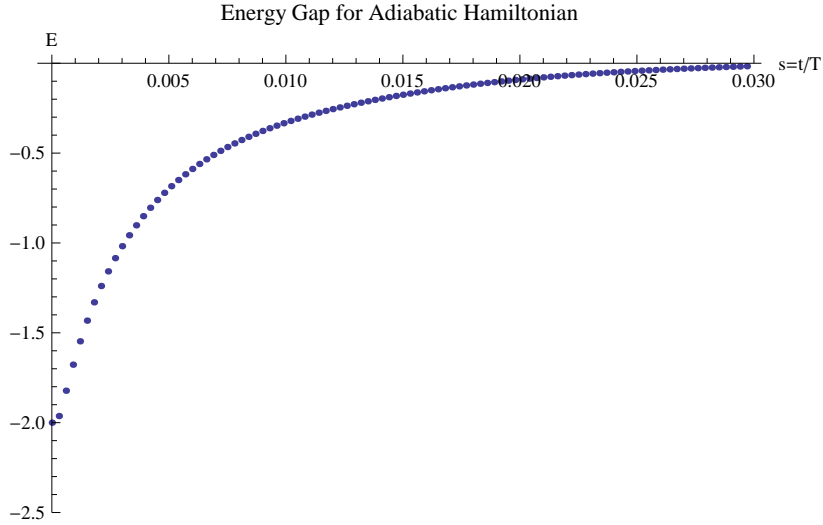} %vs .eps
\caption{Fig. 2b. Energy difference of the first and ground states energies vs $s=t/T$.}
\end{center}
\end{figure}

\pagebreak
%\medskip
\centerline{\bf Discussion of a version of quantum search}
\medskip

The concluding section of \cite{lucas} gives a problem Hamiltonian in terms of binary variables
$x_i$ for determining the largest integer in a set $n_1,n_2,\ldots,n_N$,
$$H=A\left(1-\sum_i x_i\right)^2-B\sum_i n_ix_i. \eqno(4)$$ % replaced enforces with causes:
The term with coefficient $B$ causes the largest integers to enter the sum, while this is
counterbalanced by the term with coefficient $A$ to include the sole largest.  The condition for this
Ising model to solve the problem is $A>B$ max$_i$($n_i$).  However, the quantity on the right side of
this inequality is not known a priori, and is in fact the sought-for solution.  This means that
in practice A might have to be taken arbitrarily large, and this would very likely lead to a small
gap for all initial Hamiltonians.  In turn, this implies that it is not as easy as at first glance
to recover the optimal order $O(\sqrt{N})$ of quantum search.  We recall as in the Introduction that
a modified evolution schedule for AQC is required to recover the optimal running time \cite{roland}.
Thus, it is correct that AQC may also solve computationally easy problems, but this approach %[once
%again]--[delete these 2 words later] 
does not guarantee an advantage over classical algorithms.
% write a M'ca .nb for this part and show a plot of the gap?

Figure 3a shows the gap for an evolution starting with the initial Hamiltonian $H_0=-\sum_i \sigma_x^i$
and ending with that of (4) for only three positive integers, with the largest being 6.  
% using the simple linear interpolation
% and B=1, A=(max(nvec)+3)*2
For the vast majority of the time the gap is very close to linear in $s$.  However, for small values
of $s$, as shown in Fig. 3b, there is an abrupt decrease to the global minimum before the start of the
linear growth.  For six positive integers, the largest being $91$, the behavior of the gap is very similar,
as shown in Figs. 4a and 4b.

\begin{figure}[h]
\begin{center}
\includegraphics[height=2.75in,width=4.25in,angle=0]{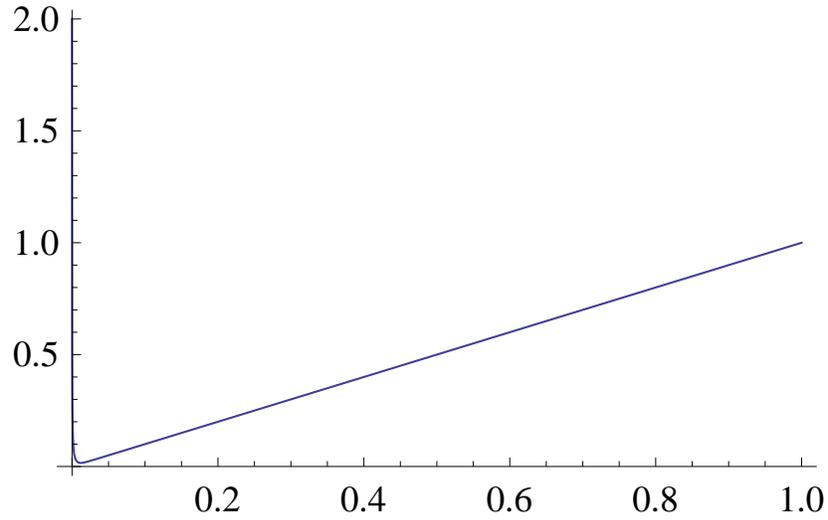} %vs .eps
\caption{Fig. 3a. Energy difference of the first and ground states energies vs $s=t/T$ for 3 input integers to be
searched.}
\end{center}
\end{figure}

\begin{figure}[h]
\begin{center}
\includegraphics[height=2.75in,width=4.25in,angle=0]{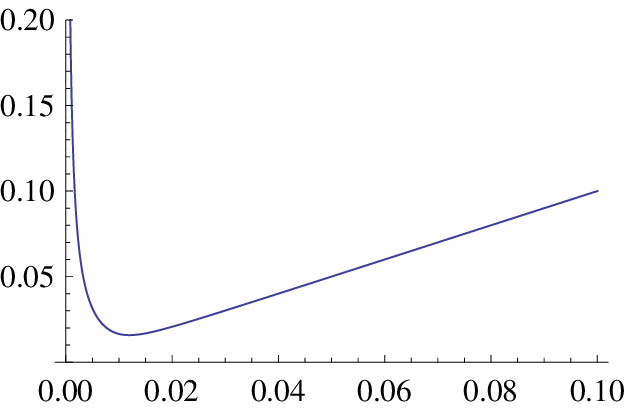} %vs .eps
\caption{Fig. 3b. Energy difference of the first and ground states energies vs $s=t/T \leq 0.1$.}
\end{center}
\end{figure}

\begin{figure}[h]
\begin{center}
\includegraphics[height=2.75in,width=4.25in,angle=0]{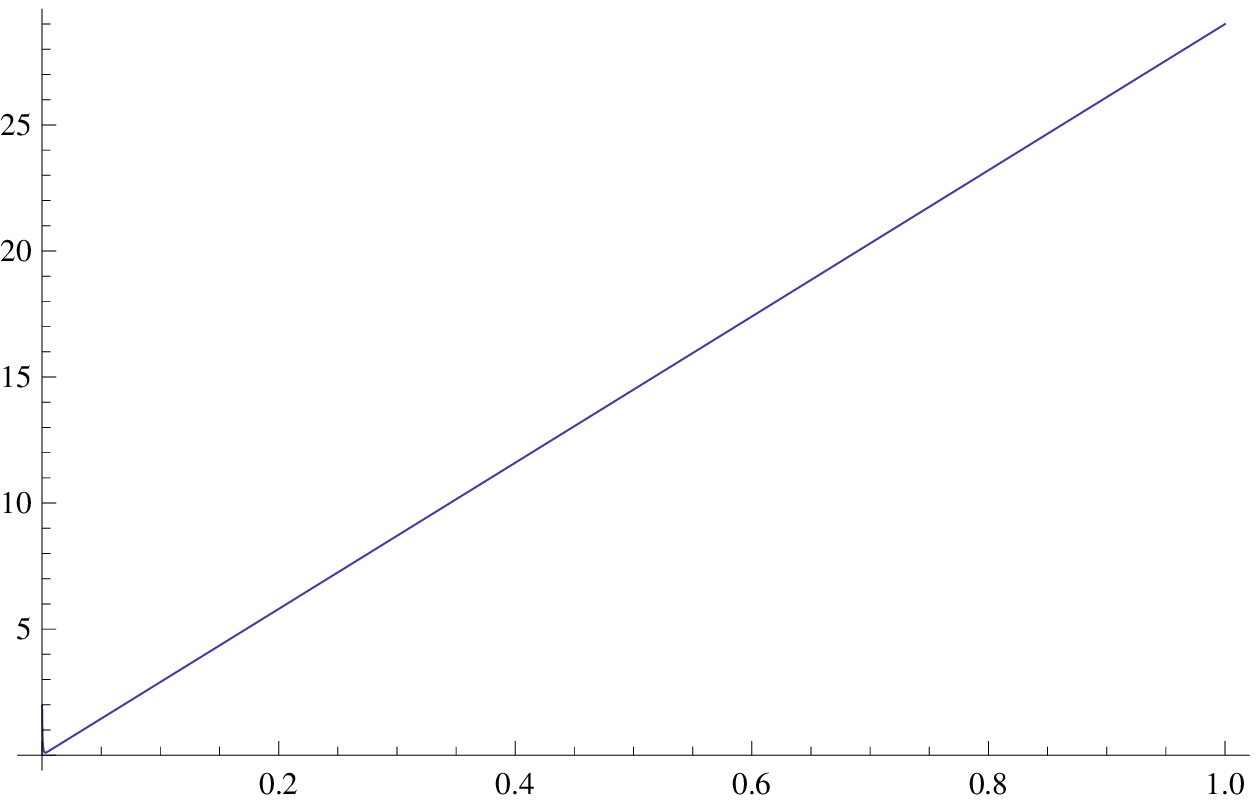} %vs .eps
\caption{Fig. 4a. Energy difference of the first and ground states energies vs $s=t/T$ for 6 input integers
to be searched.}
\end{center}
\end{figure}

\begin{figure}[h]
\begin{center}
\includegraphics[height=2.75in,width=4.25in,angle=0]{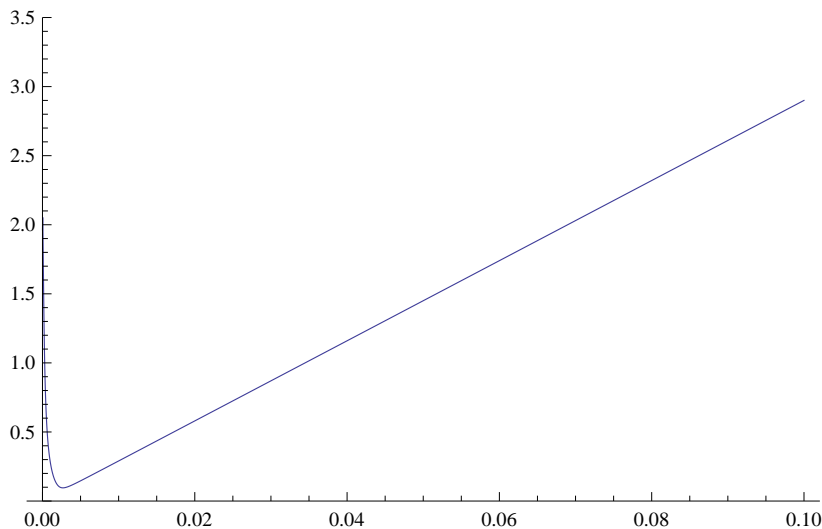} %vs .eps
\caption{Fig. 4b. Energy difference of the first and ground states energies vs $s=t/T \leq 0.1$.}
\end{center}
\end{figure}

%have an overall summary paragraph:
% would be desirable to know how the gap behaves with problem size--that could be an ambitious study

\pagebreak
%\medskip
\centerline{\bf Summary} % with an indication of future directions of research.
% with further stat. mech. analysis not made so explicit
\medskip

In summary, we have illustrated the AQC solution of the knapsack problem with both integer profits and
weights.  Indeed, it is the integrality condition of these quantities when maximizing the profit 
subject to an integer capacity $c$ that leads to the NP complexity of the problem.  
We have given specific examples using both a problem Hamiltonian requiring $n+c$ qubits for $n$ items and
a much more efficient one using $n+[\log_2 c]+1$ qubits.  Our limited numerical evidence shows that often
the spectral difference between the ground and first excited states has a minimum at small values of the normalized time $s=t/T$.  It would be of interest to know how the gap behaves with problem size.  In particular, if it could be shown that the gap decreases only polynomially (i.e., not exponentially) for larger and larger problems, then the efficiency of the AQC approach would be verified.  On the numerical side, this 
would very likely require an implementation in a compiled computer language and the running of a large number of cases, given that all of the profits, weights, and capacity are subject to variation.  
\footnote{And this is even in regard to a fixed initial Hamiltonian.} Thus further theoretical analysis is 
also of interest, which might be approached by first restricting to certain classes of knapsack problems.

Statistical mechanical analyses of the knapsack problem have been very limited.  In particular, both of the works
\cite{korutcheva} and \cite{inoue} have taken {\em all} of the profits $p_j$ to have the same constant value.
The latter article treats multiple constraints for continuous knapsack variables but the number of constraints
is directly proportional to the number of items and the investigation focuses on the capacity being one fourth of the
number of items.

We have verified the Ising model problem Hamiltonians, as expected.  Of further interest would be
alternative Hamiltonians requiring less `connectivity', i.e., fewer spin-spin couplings.  
Interestingly enough, though, there has been a recent experimental proposal for adiabatic quantum
optimization based upon ion traps, and it is thought that a variety of knapsack problems could be
solved within this framework \cite{hauke}.  This scheme is described as possible with current
trapped-ion technology by adjusting local laser intensities, in contrast to requiring specially
designed trapping potentials or a large number of laser frequencies.  In addition, it has been shown that
at the expense of using $O((n+[\log_2 c])^2)$ qubits, as a special case of an all-to-all coupled Ising model,
only local interactions in a square-lattice arrangement are required \cite{lechner}.  Thus, nearer-term experimental 
implementation may be within reach.

The subset sum problem, a very special case of knapsack, extends the number partitioning problem.
Within the Appendix we develop additional analytic expressions which serve to characterize the
transition from easy to hard instances of the subset sum problem.

\bigskip
%\pagebreak
\centerline{\bf Acknowledgements}
\medskip

The assistance of O. Orejola is gratefully acknowledged.  In turn, his support from the Colorado 
School of Mines Multicultural Engineering Program during the summer of 2016 is also.  Dr.\ P.\ Hauke
is thanked for reading the manuscript.

%\pagebreak
%\centerline{\bf Appendix A:  Higher degree traces of $H$}
%\medskip

\pagebreak
\centerline{\bf Appendix:  Relations for a subset sum problem}
\medskip

In \cite{sasamoto} the authors consider the integer solutions of $H=E$, where $H=\sum_{j=1}^N a_jn_j$.
Here the $a_j$'s are given positive integers and the $n_j$'s, each taking the values $0$ and $1$, form 
the solution(s), if they exist.  We supplement the asymptotic analysis of section 4 of \cite{sasamoto} 
which uses $W(E)$, the number of solutions of the problem $H=E$.

Let the $a_j$'s be drawn uniformly from the set $\{1,2,\ldots,L\}$ and let $\alpha=\beta/L$ be a scaled
inverse temperature.  In the limit of $N \to \infty$ with $L$ fixed, by replacing a summation with an
integral there results
$$x={E \over {NL}} \approx \int_0^1 {{ydy} \over {1+e^{\alpha y}}}.$$
The parameter describing whether at least one solution of the subset sum problem exists is
$\kappa=\log_2 L/N$.  In the limit of $N$ and $L \to \infty$, the condition $W(E)=1$ gives the critical
value
$$\kappa_c={1 \over {\ln 2}}\int_0^1 \left[\ln(1+e^{-\alpha y})+{{\alpha y} \over {1+e^{\alpha y}}}\right]dy.$$
This quantity separates the regions of hard versus easy instances of the randomized subset sum problem, with
$\kappa >\kappa_c$ being the hard-to-solve region.  

We first prove certain symmetries which are implicit in Figure 2 of section 4 of \cite{sasamoto}.
Then we make use of the dilogarithm function Li$_2(z)$ to write the functions $x=x(\alpha)$ and
$\kappa_c=\kappa_c(\alpha)$.  % then we'd really like to get x=x(\kappa_c), or the inverse fcn of that.
The latter expressions, as we indicate, provide an alternative means to show the first Proposition.

{\it Proposition 1}.  Let $-\infty < \alpha <\infty$.  Then
(a) $x(\alpha)+x(-\alpha)=1/2$, and (b) $\kappa_c(\alpha)=\kappa_c(-\alpha)$.

Hence the phase transition for this subset sum problem is determined once $x$ and $\kappa_c$ are
known for say the interval $0 \leq \alpha < \infty$.

{\it Proof}.  (a) We immediately have
$$x(\alpha)+x(-\alpha)=\int_0^1 y dy={1 \over 2}.$$
(b) We may write
$$\kappa_c(\alpha)={1 \over {\ln 2}}\left[\int_0^1 \ln(1+e^{-\alpha y})dy+\alpha x(\alpha)\right].$$
We have
$$\int_0^1 \ln\left({{1+e^{-\alpha y}} \over {1+e^{\alpha y}}}\right)dy=-\alpha \int_0^1 y dy
=-{\alpha \over 2}.$$
Then by part (a), part (b) follows since the relation $\kappa_c(\alpha)-\kappa_c(-\alpha)=0$ is
equivalent to
$${1 \over {\ln 2}}\left[\alpha(x(\alpha)+x(-\alpha))-{\alpha \over 2}\right]=0.$$ \qed

It is also of interest to separately write the functions $x(\alpha)$ and $\kappa_c(\alpha)$.  For this,
we introduce the analytically continuable dilogarithm function Li$_2(z)=\sum_{n=1}^\infty z^n/n^2$ 
($|z| \leq 1$).  In particular, Li$_2(-1)=-\pi^2/12$ and among others, there are the functional relations
$$\mbox{Li}_2\left(-{1 \over x}\right)+\mbox{Li}_2(-x)=2\mbox{Li}_2(-1)-{1 \over 2}\ln^2 x, \eqno(A.1)$$
and
$$\mbox{Li}_2(z)+\mbox{Li}_2(1-z)={\pi^2 \over 6}-\ln z \ln(1-z). \eqno(A.2)$$

{\it Proposition 2}.   Let $-\infty < \alpha <\infty$.  Then
(a)
$$x(\alpha)=-{\pi^2 \over {12\alpha^2}}+{1 \over 2}-{1 \over \alpha}\ln(1+e^\alpha)-{1 \over \alpha^2}
\mbox{Li}_2(-e^\alpha),$$
and (b)
$$\kappa_c(\alpha)=-{1 \over {\ln 2}}\left[{\alpha \over 2}+{\pi^2 \over {12\alpha}}+{1 \over \alpha}\mbox{Li}_2(-e^\alpha)-\alpha x\right].$$ % \ln\left({{1+e^{-\alpha}} \over {1+e^\alpha}}
$$=-{1 \over {\ln 2}}\left[{\pi^2 \over {6\alpha}}+{2 \over \alpha}\mbox{Li}_2(-e^\alpha)+\ln(1+e^\alpha)
\right].$$

% see my 1/5/17 pm notes:
{\it Proof}. For both parts we proceed by using power series expansion of the integrands for the
subject functions.  For (a),
$$x={E \over {NL}} \approx \int_0^1 {{ydy} \over {1+e^{\alpha y}}}=\sum_{m=0}^\infty (-1)^m \int_0^1
e^{-\alpha(m+1)y} y dy$$
$$=\sum_{m=0}^\infty {{(-1)^m} \over \alpha^2} \left[{1 \over {(m+1)^2}}-{e^{-\alpha(m+1)} \over {(m+1)^2}}
(\alpha(m+1)+1)\right]$$
$$=-{1 \over \alpha^2}\mbox{Li}_2(-1)-{1 \over \alpha^2}\sum_{m=0}^\infty (-1)^m {e^{-\alpha(m+1)} \over
{(m+1)^2}}-{1 \over \alpha}\sum_{m=0}^\infty {{(-1)^m e^{-\alpha(m+1)}} \over {m+1}}$$
$$={\pi^2 \over {12\alpha^2}}-{1 \over \alpha^2}\mbox{Li}_2(-e^{-\alpha})-{1 \over \alpha}\ln(1+e^{-\alpha}).$$

For (b) we use the integral
$$\int_0^1\ln(1+e^{-\alpha y})dy=\sum_{k=1}^\infty {{(-1)^{k-1}} \over k}\int_0^1 e^{-\alpha k y}dy$$
$$=\sum_{k=1}^\infty {{(-1)^k} \over k^2}(1-e^{-\alpha y})$$
$$={\pi^2\over {12\alpha}}+{1 \over \alpha}\mbox{Li}_2(-e^{-\alpha})$$
$$=-{\alpha \over 2}-{\pi^2\over {12\alpha}}-{1 \over \alpha}\mbox{Li}_2(-e^\alpha).$$
The latter expression follows from (A.1). \qed

An alternative proof of Proposition 2 may be based upon the integral representation
$$\mbox{Li}_2(z)=-\int_0^z {{\ln(1-t)} \over t}dt,$$
as, with a change of variable, we have
$$x(\alpha)={1 \over \alpha^2}\int_2^{1+e^\alpha} {{\ln(u-1)du} \over {u(u-1)}}={1 \over \alpha^2}\int_2^{1+e^\alpha}
\left(-{1 \over u}+{1 \over {u-1}}\right)\ln(u-1)du.$$
Upon evaluation of the integrals and the use of (A.2), the results of Proposition 2 may again be obtained.

It is now apparent how the explicit expressions of Proposition 2 may be used to verify the symmetries
of Proposition 1.  For example, for part (b) there we have, using relation (A.1),
$$\kappa_c(-\alpha)={1 \over {\ln 2}}\left[{\pi^2 \over {6\alpha}}+{2 \over \alpha}\mbox{Li}_2(-e^{-\alpha})-\ln(1+e^{-\alpha})\right]$$
$$={1 \over {\ln 2}}\left[{\pi^2 \over {6\alpha}}+{2 \over \alpha}\left[-\mbox{Li}_2(-e^\alpha)-{\pi^2 \over 6}
-{\alpha^2 \over 2}\right]-\ln(1+e^{-\alpha})\right]$$
$$={1 \over {\ln 2}}\left[-{\pi^2 \over {6\alpha}}-{2 \over \alpha}\mbox{Li}_2(-e^\alpha)-\alpha
-\ln(1+e^{-\alpha})\right]=\kappa_c(\alpha).$$
Figure 5(a) shows a plot of $x$ versus $\alpha$ and (b) a plot of $\kappa_c$ versus $\alpha$.

\begin{figure}[h]
\begin{center}
\includegraphics[height=2.75in,width=4.25in,angle=0]{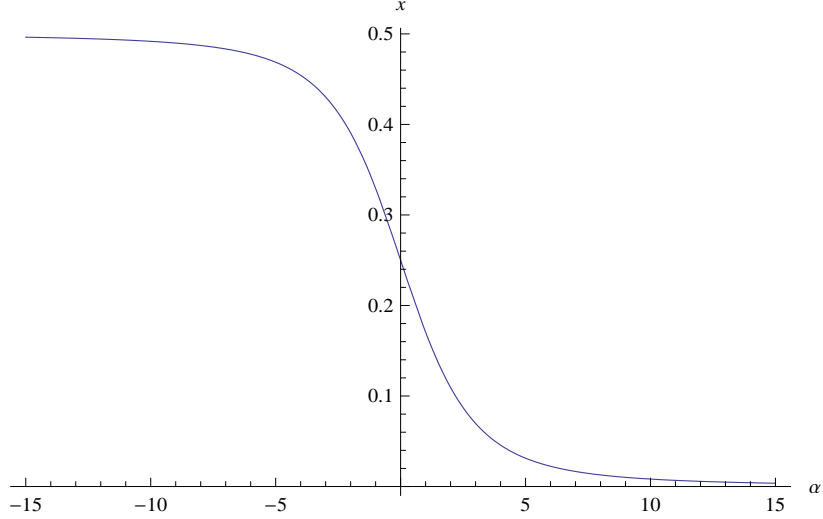} %vs .eps
\caption{Fig. 5a. The function $x=E/(NL)$ is plotted versus $\alpha=\beta/L$.}
\end{center}
\end{figure}

\begin{figure}[h]
\begin{center}
\includegraphics[height=2.75in,width=4.25in,angle=0]{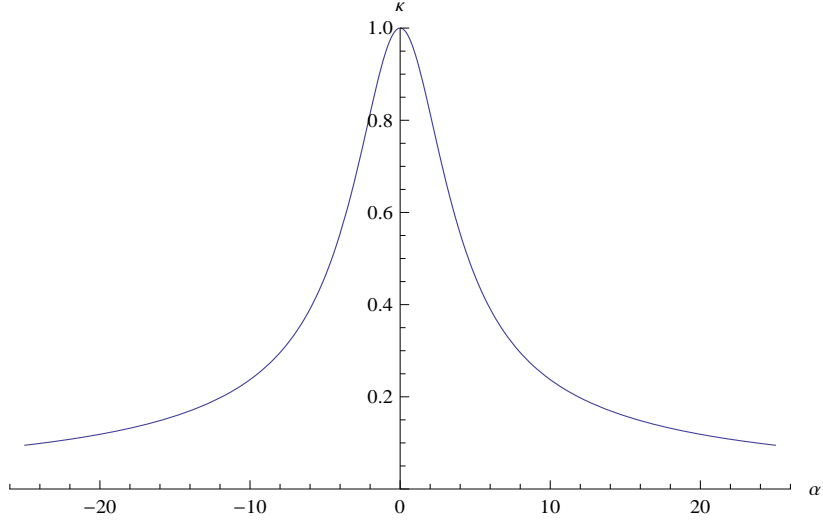} %vs .eps
\caption{Fig. 5b.  The function $\kappa_c$ is plotted versus $\alpha=\beta/L$.}
\end{center}
\end{figure}

The identification of $x$ and $\kappa_c$ in terms of the dilogarithm function is also useful in that other
points may be obtained exactly.  In particular, for $\alpha_{\mp}={1 \over 2}\ln(\mp 1+\sqrt{5})$, 
$$\mbox{Li}_2(-e^{\alpha_-})=-{\pi^2 \over {15}}+{1 \over 2}\ln^2\left({{\sqrt{5}-1} \over 2}\right),$$
and 
$$\mbox{Li}_2(-e^{\alpha_+})=-{\pi^2 \over {10}}+{1 \over 2}\ln^2\left({{\sqrt{5}+1} \over 2}\right).$$
These then give closed form evaluations of $x(\alpha_{\mp})$ and $\kappa_c(\alpha_{\mp})$.

We may now describe the curve $x$ versus $\kappa_c$ in special important regions.
\newline{\it Proposition 3}.  (a) About $x=1/4$ and $\kappa_c=1$ there holds
$$x(\kappa_c)={1 \over {12}}(3\pm 2\sqrt{6\ln 2}\sqrt{1-\kappa_c}).$$
(b) For $x$ and $\kappa_c$ near zero, there holds
$$x(\kappa_c)={{12} \over {49}}(\ln^2 2)\kappa_c^2.$$

{\it Proof}.  (a) This region is characterized by $\alpha\to 0$ and accordingly we have the expansion
$$x(\alpha)={1 \over 4}-{\alpha \over {12}}+O(\alpha^3).$$
Then $\alpha \simeq 3(1-4x)$, while
$$\kappa_c(\alpha)=1-{\alpha^2 \over {24\ln 2}}+O(\alpha^3).$$
By inverting the relation
$$\kappa_c(x) \simeq 1-{3 \over 8}{{(1-4x)^2} \over {\ln 2}}$$
we obtain the stated expression.

(b) corresponds to $\alpha \to \infty$, in which case $x(\alpha) \sim 3/(4\alpha^2)$.
This is inserted into the approximation
$$\kappa_c \approx {1 \over {\log 2}}\left[\alpha x(\alpha)+{1 \over \alpha}(1-e^{-\alpha})\right],$$
where the last term may be ignored in comparison to the others.  This gives
$$\kappa_c \approx {1 \over {\log 2}}\left({\sqrt{3} \over 2}+{2 \over \sqrt{3}}\right)\sqrt{x},$$
and rearranging provides the result.  \qed
% there could also be a final plot comparing the approximations of Prop. 3 with numerical inversion

% next approximate relations for \alpha \to 0 and \alpha \to \pm \infty.
% then also add remarks to the Summary relating to this Appendix? ......

There are differential relations between the scaled energy and the critical value $\kappa_c$.  For example,
we have the following.
\newline{\it Proposition 4}.  There holds
$$\kappa_c(\alpha)={1 \over {\ln 2}}\int\alpha dx(\alpha).$$

{\it Proof}.  We have
$$x(\alpha)=-{d \over {d\alpha}}\int_0^1 \ln(1+e^{-\alpha y})dy,$$
yielding
$$x(\alpha)=-{d \over {d\alpha}}[(\ln 2)\kappa_c(\alpha)-\alpha x(\alpha)].$$
Then 
$$\alpha {{dx(\alpha)} \over {d\alpha}}=(\ln 2){{d\kappa_c(\alpha)} \over {d\alpha}},$$
giving the result.  \qed

Therefore, from $dx/d\kappa_c=(\ln 2)/\alpha$ we know that the $x-\kappa_c$ curve has negative slope for $\alpha<0$
and positive slope for $\alpha>0$.

We may also mention the constrained subset sum problem, for which the number of chosen $a_j$'s is fixed to
an integer $M>0$, so that now $\sum_{j=1}^N n_j=M$.  In this setting, the statistical mechanical analysis
uses the grand canonical ensemble \cite{sasamoto2} with chemical potential $\mu>0$.  

% to give:  x(\alpha,\mu), part of a Proposition with $x(\alpha,\mu)+x(\alpha(-\alpha,-\mu)=1/2$,
% \rho(\alpha,\mu), x(\alpha,\mu) explicitly in terms of the dilog function; but D(\alpha,\mu) & W omitted.
% and at least also \kappa_c(\alpha,\mu)

We introduce the following quantities \footnote{With $\alpha=\beta/L$ as before, despite the first line of p.\
371 of \cite{sasamoto2}.}
$$\rho={M \over N}\approx \int_0^1 {{dy} \over {1+e^{\alpha y-\mu}}}, ~~~~
x={E \over {NL}}\approx \int_0^1 {{ydy} \over {1+e^{\alpha y-\mu}}},$$
and the critical value
$$\kappa_c(\alpha)={1 \over {\ln 2}}\int_0^1 \left[\ln(1+e^{-\alpha y+\mu})+{{\alpha y} \over {1+e^{\alpha y-\mu}}}
-{\mu \over {1+e^{\alpha y-\mu}}}\right]dy.$$

We now have the following mathematical symmetries:
$$\rho(-\alpha,\mu)+\rho(\alpha,\mu)=1,$$
$$x(\alpha,\mu)+x(-\alpha,-\mu)={1 \over 2},$$
and
$$\kappa_c(\alpha,\mu)=\kappa_c(-\alpha,-\mu).$$
We omit proofs of these relations as well as of explicit expressions which we supply next.

For $\rho=M/N$, \footnote{Note accordingly that a correction for an exponent in the expression for $\mu=\mu(\alpha,
\rho)$ on p.\ 371 of \cite{sasamoto2} is needed.}
$$\rho=1-{1 \over \alpha}\ln\left({{1+e^{-\mu}} \over {1+e^{\alpha-\mu}}}\right).$$
For the scaled energy,
$$x(\alpha,\mu)={1 \over {2\alpha}}\left[\alpha-2\ln(1+e^{\alpha-\mu})\right]+{1 \over \alpha^2}\left[\mbox{Li}_2
(-e^{-\mu})-\mbox{Li}_2(-e^{\alpha-\mu})\right].$$
For the critical value $\kappa_c$, for which $\kappa>\kappa_c$ is a region of hard problem instances for the
constrained subset sum problem,
$$\kappa_c(\alpha,\mu)={1 \over {\ln 2}}\left\{-\ln(1+e^{\alpha-\mu})+{2 \over \alpha}\left[\mbox{Li}_2
(-e^{-\mu})-\mbox{Li}_2(-e^{\alpha-\mu})\right]\right\}.$$
We also omit expressions for the number of solutions $W(E,M)$ of the constrained problem.  These may be written in
terms of the logarithm, dilogarithm, and trilogarithm functions.  The reduction of the expressions for $x(\alpha,\mu)$
and $\kappa_c(\alpha,\mu)$ as $\mu \to 0$ for the unconstrained case is obvious.

\pagebreak

\end{document}